\documentstyle[sprocl]{article}
\input{psfig}
\psrotatefirst 
\def\Journal#1#2#3#4{{#1} {\bf #2}, #3 (#4)}

\def\PRD{{\em Phys. Rev.} D}

\def\be{\begin{equation}}
\def\ee{\end{equation}}
\def\bea{\begin{eqnarray}}
\def\eea{\end{eqnarray}}

\begin{document}
\title{SEARCH FOR T-VIOLATION in $K_{\mu 3}$ DECAY}
\author{
M.V.~Diwan, J.~Frank, A.~Gordeev, S.~Kettell, L.~Leipuner, 
L.~Littenberg, \underline{H.~Ma}, V.~Polychronakos }
\address{
Brookhaven National Laboratory, Upton, NY}
\author{ 
G.~Atoyan, V.~Issakov, O.~Karavichev, S.~Laptev, A.~Poblaguev, 
A.~Proskuryakov }
\address{
Institute for Nuclear Research, Moscow, Russia}
\author{B.~Barakat, M.~Elaasar, D.~Greenwood, K.~Johnston }
\address{Louisiana Tech University, Ruston, LA}
\author{R.~Adair , R.~Larsen}
\address{Yale University, New Haven, CT} 

\maketitle\abstracts{
We propose a  new experiment at the AGS to 
search for the T-violating polarization of
the muon normal to the decay plane of the $K^+ \to \mu^+ \pi^0 \nu$ decay.
Motivated by the need for a stronger CP violation source to account
for the baryon asymmetry of the Universe, the experiment aims to
search for T-violation beyond the Standard Model. 
 The experiment  will be performed with in-flight decays from
 an intense 
 2 GeV/c separated $K^+$ beam 
 at the AGS.
 We expect to analyze  $10^9$ events to
 obtain the sensitivity  of $\delta P_t = \pm 0.00013$
 at 1 $\sigma$, corresponding to the sensitivity of $\pm 0.0007$ to
$Im\xi$,
an improvement by 40 over the present limit.
}

\section{Introduction}
We propose a new search for the time reversal violating
polarization of the muon normal to the  plane of
the $K^+ \to \mu^+ \pi^0 \nu$ decay~\cite{proposal}.  
The term $\vec{\sigma_\mu}
\cdot (\vec{P_\pi} \times \vec{P_\mu})$,
which is proportional to the projection of the muon polarization out
of the
decay plane,
changes sign upon time reversal; therefore a finite expectation value
for this quantity indicates a violation of time reversal invariance.
Moreover, since the Standard Model prediction for such polarization is zero, 
and there is no final state interaction, 
the observation of T-violation in the $K_{\mu 3}$ decay is a discovery of
T-violation beyond the Standard Model. 
Through the CPT theorem we know that T-invariance is intimately
related to CP-invariance.  
Although the only observed CP-violation in the
neutral kaon system can be attributed to the complex phase in the 
CKM matrix within the Standard Model 
  the true nature of CP-violation is far from
being revealed by the current experimental data. 
It is now accepted that the baryon asymmetry of the universe 
requires a source of CP violation stronger than that embodied in the
CKM matrix.
Models of non-standard CP violation that produce the baryon asymmetry
could also produce effects observable in the transverse polarization.
Because of  the very high sensitivity
of the experiment the possibility of discovering unexpected new
physics should not be underestimated.

 The best previous experimental limits were obtained  over 15 years
ago with 4 GeV charged kaons~\cite{campbell} at the
AGS, yielding a result of $P^T_\mu = 0.0031 \pm 0.0053$.
The high intensity kaon beams available now at the AGS
makes it possible to improve the limit on the polarization
by more than an order of magnitude.

\section{Detector Design}
\begin{figure}
\psfig{figure=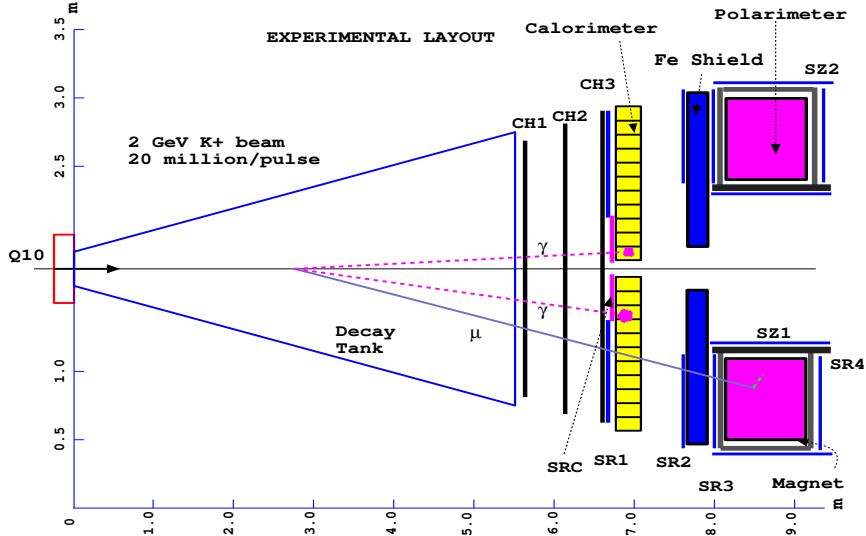,height=2.8in,width=4.5in,angle=90} 
\caption{Schematic of the detector.  A typical $K^+\rightarrow
\mu^+\pi^0\nu$ events is superimposed.}
\label{pict1}
\end{figure}
The experiment will be performed with 2 GeV/c electrostatically
seperated charged kaons
decaying in flight. 
The beam intensity will be $2\times 10^7 K^+$'s/spill with 
$3\times 10^{13}$ protons on target every 3.6 sec. 
Figure \ref{pict1}
 shows the plan  view
 of the experiment.  The basic workings of the experiment
 are the same as the experiment in Reference 2.
 The detailed design is, however, optimized for a
 high intensity 2 GeV beam. 
The cylindrically symmetric detector
is centered on the kaon beam. The $K^+_{\mu 3}$ ~decays of interest
occur in the decay tank;  the photons from the
$\pi^0$ decay are detected in the calorimeter; the muon stops
in the polarimeter.
The decay of the stopped muon is 
detected in the polarimeter by wire chambers, which are
arranged radially with graphite wedges that serve as absorber medium.
The hit pattern in the polarimeter identifies the muon stop as well as 
positron direction relative to the muon stop.
By selecting events with $\pi^0$ moving along the beam direction and
muon moving perpendicular to the beam direction in the $K^+$ center of  
mass frame, the decay plane coincides with the radial wedges. 
A  non-zero transverse muon polarization causes an asymmetry between
the number of muons that decay
clockwise versus the number counter-clockwise.
To reduce systematic errors, a weak solenoidal magnetic field along
the beam direction 
(70 gauss or an precessing period of $\sim 1 \mu s$) with
polarity reversal every spill is applied to the polarimeter. 
The initial muon transverse polarization causes a small shift in the
phase of the sinusoidal oscillation in the measured asymmetry. 
The difference in the asymmetry for the two polarities is proportional 
to the muon polarization in the decay plane, while the sum is
proportional to the muon polarization normal to the decay plane. 

Compared to the previous experiment,  this experiment has
much better background rejection and event reconstruction.  
The separated $K^+$ beam should greatly reduce the accidental rate.
The polarimeter is fine segmented and the analyzing power is higher. 
The positron signature is defined by the coincidence of signals
in a pair of neighboring wedges.
The larger calorimeter makes it possible to reconstruct the $\pi^0$ 
momentum. Together with the muon trajectory, the event can be fully
reconstructed.  
The detector
acceptance and background rejection is optimized using GEANT
simulation. 
We expect to obtain 550 events/spill, with up to 20\% 
background.  With an analyzing power of over 30\%, we expect to 
reach the statistical sensitivity of 
$\delta P_t = 1.3\times 10^{-4}$.

	At such a high statistical accuracy, much care has to be taken 
in reducing systematic errors.  We have studied various effects that
may give a false signal, such as misalignment within the polarimeter, 
misalignment among and detector components and the beam, asymmetry in
or caused by the precessing field and inefficiencies. 
We believe that 
these errors can be made acceptably small by proper construction techniques 
and by using symmetries of the apparatus to internally cancel the
systematic errors.  In addition, we will use the T-conserving 
component of the muon polarization to calibrate the detector analyzing 
power, and samples of muon stops with no known transverse
polarization,  such as muons from $K_{\mu 2}$ decays, to detect any 
systematic bias. 

	The proposal has been submitted to the Laboratory in Aug 1996.
If approved and funded, we would like to have the first engineering
run in  1998.  The physics data taking will take about 2000 hours 
of running time.

\end{document}